\documentclass[authoryear,12pt,3p]{jowarticle}

\usepackage[english]{babel}
\usepackage[utf8]{inputenc}
\usepackage{amsmath}
\usepackage{graphicx}
\usepackage[colorinlistoftodos]{todonotes}
\usepackage{natbib}

\makeatletter
\renewcommand\section{\@startsection{section}{1}{\z@}{-3.25ex plus -1ex minus -.2ex}{1.5ex plus .2ex}{\normalsize\bf}}
\renewcommand\subsection{\@startsection{subsection}{2}{\z@}{-3.25ex plus -1ex minus -.2ex}{1.5ex plus .2ex}{\normalsize\bf}}
\renewcommand\subsubsection{\@startsection{subsubsection}{3}{\z@}{-3.25ex plus -1ex minus -.2ex}{1.5ex plus .2ex}{\normalsize\bf}}
\makeatother

\begin{document}
\begin{frontmatter}
\title{Scientific Polarization}

\author{Cailin O'Connor}\ead{cailino@uci.edu} \author{James Owen Weatherall}\ead{weatherj@uci.edu}
\address{Department of Logic and Philosophy of Science \\ University of California, Irvine}

\date{\today}

\begin{abstract}
Contemporary societies are often “polarized”, in the sense that sub-groups within these societies hold stably opposing beliefs, even when there is a fact of the matter.  Extant models of polarization do not capture the idea that some beliefs are true and others false.  Here we present a model, based on the network epistemology framework of Bala and Goyal ["Learning from neighbors", \textit{Rev. Econ. Stud.} \textbf{65}(3), 784-811 (1998)], in which polarization emerges even though agents gather evidence about their beliefs, and true belief yields a pay-off advantage.  As we discuss, these results are especially relevant to polarization in scientific communities, for these reasons.  The key mechanism that generates polarization involves treating evidence generated by other agents as uncertain when their beliefs are relatively different from one's own.
\end{abstract}
\end{frontmatter}

\section{Introduction}

Is anthropogenic climate change real?  This question, asked in the wrong setting, will spark a furious debate.  Some members of the U.S. public are convinced that global warming is a liberal conspiracy dreamt up to restrict personal liberties.  Others believe that climate change is the most serious existential threat facing humanity.  Although there has long been a consensus among climate scientists that anthropogenic warming poses serious risks \citep{OreskesClimate}, there does not appear to be an emerging consensus concerning this issue among the American public at large \citep{McCright+Dunlap}.  This is an example of what is sometimes called ``polarization''---subgroups within a society maintain stable, opposing beliefs, even in the face of extensive debate on an issue.\footnote{Some authors use the term ``polarization'', or more specifically, ``belief'' or ``attitude polarization'', to refer to the more limited phenomenon in which two individuals with opposing credences both strengthen their beliefs in light of identical evidence.  Other authors, particularly in psychology, use ``group polarization'' to refer to situations in which discussion among like-minded individuals strengthens individual beliefs beyond what anyone in the group started with.  As noted, we are using the term ``polarization'' in a sense common in political discourse, to describe situations in which beliefs or opinions of a group fail to converge towards a consensus, or else actually diverge, over time. \citet{bramson2017understanding} differentiate between ways one might define or measure polarization in this more general sense.}

There is now a large literature that attempts to model polarization in the socio-political realm.\footnote{We survey this literature in section \ref{sec:modelpolar}; see  \citep{bramson2017understanding} for a review.} A general take-away from this body of work is that polarization can occur when agents influence each others' opinions, but where the degree of this influence depends on the similarity between agents' opinions.  This sort of situation can generate feedback loops that stabilize polarization. Subgroups form where actors share beliefs and, as a result, are only influenced by those in their group.

The models considered in this literature do not generally treat beliefs as having different truth-values, and agents do not influence one another by sharing evidence supporting their beliefs.  In other words these models show how polarization can emerge from various opinion dynamics, but not how polarization can persist in the face of evidence demonstrating that acting in accordance with one belief yields a distinct advantage.\footnote{As we discuss below, there are some exceptions to this generalization---most notably, in work by \citet{olsson2013bayesian}---but the model we present here is substantially different and, we believe, more perspicuous.}  In many cases of polarization, this seems appropriate, since the underlying positions of those involved are motivated by moral, religious, or political values.  On abortion, for instance---another issue on which the U.S. public is polarized---religious beliefs motivate many of those who oppose legalized abortion, while those who support it are often driven by feminist values.  Similarly, social and political values play a role in attitudes about climate change:  liberals tend to believe that central governing bodies have a responsibility to protect shared environmental resources, while conservatives argue that a free market will generate suitable responses endogenously, without government intervention.

But differences in values of this sort are not the only important aspect of polarization.  For instance, in the case of climate change, it is not merely that there are disagreements concerning what policies to adopt; there is also polarization in belief concerning matters of fact about the causes and likely consequences of global warming.  This is so despite the fact that there is ample evidence available and the long-term consequence of injudicious policies are potentially severe.  Indeed, polarization can appear even in communities that broadly share values---including in scientific communities, which can become deeply divided over issues such as what foundational theory to adopt, what methodology is appropriate, or what the truth of the matter is in some case.  For instance, as we will discuss below, researchers working on Lyme disease seem to have polarized.  How can agents acting under such conditions reach opinions that are so deeply divided?

Our aim in this paper is to show how a group of learners who share evidence, and who have the same aims and values, can nonetheless become polarized.  We do so by presenting a simple model, based on the network epistemology framework developed by economists \citet{venkatesh1998learning} and introduced to philosophy of science by \citet{zollman2007communication}, in which agents gather and share evidence concerning which of two possible actions yields a better expected payoff.  In this model, all agents have the same preferences and there is a fact of the matter concerning which action is preferable.  The agents all have access to the same evidence, which they continually gather and use to update their beliefs.  However, they treat evidence from other agents as uncertain, using a simple heuristic according to which agents whose beliefs are distant from their own are judged to be more epistemically unreliable.  As we show, epistemic communities in which agents employ this heuristic can become stably polarized.  As a result of this polarization, the accuracy of scientific beliefs in the community is typically worse.

Furthermore, while we do not claim that employing this heuristic is individually rational, we do claim that it is justifiable and that similar heuristics are widespread.  Indeed, it is essential to scientific practice that scientists make judgments concerning the reliability of other scientists' work, and condition their beliefs accordingly.\footnote{Psychologists often appeal to motivated reasoning in explaining polarization.  For example, humans tend to engage in confirmation bias, which involves seeking out and assimilating new information supporting their already deeply held beliefs \citep{lord1979biased}.  But that is not what is going on here: agents to not selectively update on evidence that is probable given their current beliefs; they do so on the basis of their judgments about the \emph{source} of the evidence, irrespective of what the evidence tends to support.  We take this to be more epistemically justifiable---which makes the appearance of polarization in the presence of this heuristic more surprising.}  That this sort of judgment can, at least in some cases, lead to polarization is therefore striking---and can help explain why we observe polarization in real scientific communities.  It also provides a novel example of how individually justifiable epistemic heuristics can lead to group-level behavior that is not truth-conducive \citep{mayo2011independence}.

In the next section, we present the case of Lyme disease, wherein a scientific community has become highly polarized.  Section \ref{sec:modelpolar} will describe other models of polarization and then introduce the model we will analyze here.  In section \ref{sec:results} we describe the main results of the paper, which are that treating the evidence produced by those with whom we disagree as uncertain can lead to polarization, and that this impairs the ability of a scientific community to achieve true beliefs.  In the conclusion, we discuss implications of the work presented here, both for philosophy of science and for social epistemology.

\section{Chronic Lyme and the Polarization of Science}

Rheumatologist Allen Steere first identified Lyme disease as a new, tick-borne illness during the mid-1970s.  At the time, hundreds of people in Lyme, Connecticut and surrounding communities were suffering from a mysterious set of maladies---joint pain, arthritis, extreme fatigue, headaches, brain fog.  One of these sufferers, Polly Murray, was referred to Dr. Steere.\footnote{This history was reported in the New York Times article ‘Stalking Dr. Steere over Lyme Disease’, published June 17, 2001.}  Diligent work by Steere and others eventually linked these symptoms to tick bites, and not long thereafter the spirochete responsible was isolated by medical entomologist Willy Burgdorfer, and named \textit{Borrelia burgdorferi} is his honor \citep{steere1977epidemic,burgdorfer1982lyme}.

This discovery was a savior for patients like Polly Murray and others infected with Lyme.  Since the spirochete is treatable with antibiotics, a course of therapy was often enough to make a drastic difference in the lives of those who had been infected.  Despite this, by the 1990s, Steere was receiving death threats from angry Lyme patients.  How did the man whose discoveries paved the way for everything we know about Lyme end up a reviled figure by the very patients he sought to cure?  In the early 1990s, Steere became worried that Lyme was being treated as catchall for patients with Lyme-like symptoms who could not be otherwise diagnosed.  He also worried that harmful courses of antibiotics were being prescribed to patients who did not need them.  After investigation, he began to advocate for more careful diagnosis and treatment of Lyme \citep{steere1993overdiagnosis}.  Thus began the ``Lyme Wars''.  At the heart of this now decades long scientific debate are 1) the question of whether Lyme can persist in patients after a short cycle of antibiotics, and 2) the question of whether long term doses of antibiotics are successful in improving the symptoms of Lyme patients.

On one side are thousands of patients, and the physicians who treat them, who say that ``chronic Lyme'' is ruining their lives.  They describe debilitating symptoms similar to those known to occur if Lyme goes untreated and enters the late stage of the illness---arthritis, pain, fatigue, and a host of cognitive problems.  Many of them seek treatment from ``Lyme-literate'' physicians, who claim that intravenous, long term courses of antibiotics are both necessary and successful in treating these patients.  As they point out, studies have shown that even after intense courses of antibiotics, macaques and some other species can still test positive for Lyme and can even reinfect ticks with the Lyme spirochete \citep{embers2012persistence,straubinger2000status}.  Documentaries, such as the 2008 \textit{Under Our Skin}, and first hand accounts such as Allie Cashel's 2015 \textit{Suffering in Silence} describe the horrors of chronic Lyme, and portray the doctors who do not believe in it as either incompetent or under the sway of insurance firms who do not want to pay for long courses of treatment.

On the other side of the debate are the majority of physicians, including Steere, who believe that chronic Lyme is actually a combination of post-Lyme syndrome--a set of symptoms that are the result of previous damage by \textit{Borrelia burgdorferi} rather than current infection---and other diseases, such as fibromyalgia and chronic fatigue syndrome, that are themselves poorly understood \citep{steere2004emergence}.  They point out that random control trials have shown no benefit for long term antibiotic treatment for chronic Lyme patients \citep{klempner2001two}, and that, in many cases, those who are sick do not test positive for Lyme.

In the case of chronic Lyme, researchers have apparently failed to approach consensus, and even have even become increasingly convinced that those on the other side are not to be trusted.  In other words, they are polarizing in much the way the public sometimes polarizes on political and social issues.  The surprising thing about this case, though, is that many values and goals are the same on both sides of the debate.  By this we mean that both establishment physicians and Lyme literate ones seem to want to reduce the suffering of Lyme patients.\footnote{It is, of course, possible that there are some Lyme researchers influenced by industry funding, or who are trying to bilk patients.  This does not seem to be the case for most of the physicians involved.}  They all seem to want to discover the truth of what is happening with chronic Lyme.  They all have access to similar sorts of evidence---they see and treat patients with Lyme and they read the same articles.  In a case like this, we might expect beliefs in the group to start converging to a consensus.  The fact that this does not seem to be happening presents a puzzle.

One striking feature of the Lyme disease research community, as noted, is profound mistrust between groups adhering to different views concerning the status of chronic Lyme disease.  Members of these two groups have access to the evidence gathered by the other, but they appear to discount it.  Establishment physicians think Lyme literate physicians are quacks, and so put little weight on their clinical experience and, arguably more rigorous, published studies.  Lyme literate physicians think that Steere and his collaborators are swayed by industry interests, and are producing biased science.  As we will describe in the next section, this sort of relationship between belief and trust is what often generates polarized opinions in models of polarization.  What is novel here is the observation that we can get these sorts of polarized outcomes in a scientific community where researchers are explicitly motivated by epistemic aims, where they gather evidence from the world, and where they use reasonable heuristics in determining how to update their beliefs in light of new evidence.

\section{Modeling Polarization}
\label{sec:modelpolar}

Empirical work suggests that, in general, discussion tends to lead to greater agreement among individuals.\footnote{For example, \citet{festinger1950social}, in a classic study, showed how location around housing courts, and thus social interaction, importantly determined opinions in a study of MIT students.  Students tended to adopt the beliefs of their neighbors.}  Many models of influence between people attempt to capture and explain this tendency towards shared opinions \citep{axelrod1997dissemination,hegselmann2002opinion}.  The question for those interested in polarization is why, despite this general tendency, some groups never converge to uniform beliefs.  The key ingredient to generating polarization in models of opinions dynamics is to make the social influence of one individual on another dependent on how similar their beliefs or positions are. Many models have been developed in which incorporating this sort of dependence can produce polarization.\footnote{Again, for a philosophically sensitive review of models of polarization see \citep{bramson2017understanding}.}

In an influential example, \citet{hegselmann2002opinion} assume that individuals in a group hold some opinion between 0 and 1.\footnote{In work that predates this, \citet{axelrod1997dissemination} provides a model where cultures are represented by variants (lists of numbers) and where similarity of these variants determines how likely they are to adopt other variants from neighbors in a grid.  In this way `cultural similarity' determines cultural influence.  As he shows, stably different cultures, which we might think of as polarized in some sense, can co-exist if they have no overlap and thus do not influence each other at all.}  These individuals update their opinions over time by averaging them with group members, but they only include group members whose current beliefs are within some distance of their own.  As they show, as this distance grows smaller, groups will fail to reach consensus on an opinion, and instead form subgroups, each of which jointly holds the same opinion.\footnote{They label the outcome where just two subgroups with divergent opinions emerge as `polarization'.}  If similarity of belief is not taken into account in determining influence, on the other hand, the group always converges to consensus.\footnote{In this tradition, see also \citet{deffuant2002can,deffuant2006comparing}.}   \citet{macy2003polarization} look at networked agents who adopt binary opinion states, and whose states are influenced by neighbors depending on weights they assign to them.  The weights in turn update based on the similarity of opinion states, leading to polarized groups who do not trust each other.  \citet{baldassarri2007dynamics} assume that individuals only interact with those who have similar interests and similar opinions, and observe both polarization of beliefs and homophily---where two groups stop interacting with each other based on their different beliefs.\footnote{See also \citet{galam1991towards,galam2010public,galam2011collective,nowak1990private,mas2013differentiation,la2014influence}.  In addition, a number of modelers have shown how belief polarization---updating in different directions for the same evidence---can be rational.  This can occur under the right conditions for agents with different priors or with different background beliefs \citep{dixit2007political, jern2014belief,benoit2014theory}.}

The models just described all concern cases in which actors choose between opinions or beliefs that are equally good, in the sense that there are no external reasons to hold one belief over another.  Agents do not seek evidence from the world in forming their beliefs, and beliefs play no role in action.  There are a few models in the literature which look instead at cases where one belief is superior.  \citet{hegselmann2006truth} give a model much like their original opinion model, but where some or all agents are attracted to one `true' opinion, while also taking peer opinions into account.  They take this attraction to correspond to an ability on the part of the agents to gather evidence or observe the world in such a way that guides their opinions toward truth. They find, however, that whenever each individual has even a small attraction to the truth, they all eventually find it.  In other words, polarization is only possible if at least some agents care only about the opinions of others.  In this sense, their model does not seem to capture what is going on in cases of scientific polarization.\footnote{For more work in this framework see \citet{kurz2011hegselmann,liu2014multi}.  One key difference between these models and ours is that their agents can never come to disregard, or give up on, a possibly true theory.}

The other two models in this realm aim at understanding group deliberation.  \citet{Singer2017} look at a deliberating group that shares ``reasons'' for belief---positive and negative numbers which they add up to draw a conclusion---chosen from some fixed set. They show how subgroups can polarize when individuals only forget those reasons that do not cohere with their most likely current conclusion.  In this model, however, the `'`reasons'' do not correspond to evidence gathered from actually performing actions informed by the belief. In addition, while their model arguably has a representation of a better position (i.e., the one with more reasons to support it), they do not represent agents capable of having true or false beliefs as we do.

The model most like ours is from \citet{olsson2013bayesian}, who use the Laputa network epistemology framework to investigate polarization.\footnote{This framework is first presented in \citet{angere2010knowledge}.}  In their model, agents deliberate over a proposition $p$, and test the world with varying degrees of accuracy.  They use Bayesian updating to adjust their credences in $p$ based on their evidence and others' statements of belief.  They can adjust their levels of trust in others' statements (and thus their updating) based on whether they have similar beliefs.  The key difference between the models is that their agents communicate by stating their beliefs.  Ours actually share unbiased evidence with each other---and yet, as we will see, polarization still appears.\footnote{In addition, the way their agents gather evidence arguably less closely mimics many cases of scientific process as they receive private signals from a distribution, rather than sampling data points.}

\subsection{Epistemic Networks and Uncertain Evidence}

As noted above, we work in the network epistemology framework developed by \citet{venkatesh1998learning}.  In this model, agents decide between two actions that have different probabilities of yielding some fixed payoff.  The agents choose which action to take based on their current belief about which has the highest expected return.  That belief, in turn, is informed by the success of their own past actions and those of other agents as shared in a social network.  \citet{zollman2007communication} adapts this model to represent the emergence of scientific consensus.\footnote{This framework has been used in philosophy of science by \citet{zollman2010epistemic,mayo2011independence,kummerfeld2015conservatism,holman2015problem,rosenstock2017epistemic,Weatherall+etal,OConnor+WeatherallConformity,OConnor+WeatherallBook}. \citet{zollman2013network} provides a review of the literature up to 2013.}

In more detail, the basic model consists of a network of agents, each of whom is connected to some or all of the other agents in the network.  The agents decide between two actions: action A (All right) and action B (Better). Action A is well understood, and all agents know that performing it generates success with probability .5.  The success rate of action B is uncertain: agents know that action B is either slightly better (success rate of $.5+\epsilon$) or slightly worse (success rate of $.5-\epsilon$), but they do not know which case obtains.  In fact, action B has a success rate of $.5+ \epsilon$, and so action B is preferable to action A.  The goal is for agents to determine which of the actions has a higher success rate.  This is an example of a ``two-armed bandit problem'', so called because it matches a case where actors must choose one of two arms on a slot machine that yield payoffs with differential rates.\footnote{ The version of the model we consider here follows \citet{zollman2007communication} very closely, because unlike later versions considered by \citet{zollman2010epistemic} and others, the beliefs of the agents in the 2007 model are captured by a single number.  This made representing distance between agents beliefs in our model much more tractable.}

Each agent in the network has some credence between 0 and 1 that action B is better than action A.  An agent with credence .54, for example, thinks there is a 54\% chance that B is the better action.  These credences are initially randomly assigned.  In each round of simulation, actors choose the action that they believe has the highest expected outcome.  If their credence is $<.5$ they choose action A, otherwise action B.  (In what follows, we sometimes say that an agent ``accepts theory A'' if their credence is $<.5$, and thus they believe action A is better; otherwise they ``accept theory B''.)  They then perform their action some (fixed) number of times, $n$, and observe how often it succeeds.\footnote{Note, this parameter was added to the model by \citet{zollman2007communication} and does not appear in the work of \citet{venkatesh1998learning}.}  Agents subsequently use Bayes' rule to update their credence based on both their own experience, and the experiences of their neighbors  on the network.  Since there is no uncertainty regarding action A (it is known to succeed exactly half the time), performing action A provides no information to the agents; thus, agents' beliefs change only if they or at least one of their neighbors test theory B.\footnote{Notice that this framework can model situations outside the realm of science.  The main features of interest are agents who choose between two actions, have beliefs about the efficacy of these actions, and share evidence relevant to these beliefs.  We take these to be key features of scientific communities, but also some other sorts of everyday communities where individuals share evidence relevant to belief.}

As agents update their beliefs over time, one of two things happens.  Either all agents come to (erroneously) accept theory A, and so do not gather new, informative evidence, or else all agents come to accept theory B with very high credence, so that the chances they ever revert to incorrect beliefs become vanishingly small.  In other words, the network tends to arrive at a consensus, wherein all agents (approximately) stably accept either the true theory or the false theory.

In the Bala-Goyal model as we have been describing it, all agents treat evidence gathered by themselves and other agents in the network in the same way.  But as we saw in the case of Lyme disease discussed above, under some circumstances, members of a scientific community stop trusting the evidence produced by other colleagues.  In such cases, agents do not update their beliefs on evidence produced by other agents in the way that they do on the evidence they produce themselves.  To capture the dynamics of such a situation, we alter the model studied by \citet{venkatesh1998learning} and by \citet{zollman2007communication} so that agents treat evidence produced by other agents as uncertain.  These agents then update their beliefs using Jeffrey's rule instead of Bayes' rule.

Jeffrey conditionalization, unlike strict Bayesian conditionalization, allows actors to update beliefs in light of uncertain  evidence.\footnote{See \citet[Ch. 11]{Jeffrey}.}  We use it in the present case as follows.  Suppose that Ian tells Jill he observed some evidence, $E$.  Further suppose that Jill does not fully trust Ian's data gathering practices, meaning she has credence $P_f(E) \leq 1$ that the evidence he described obtained.  Under Jeffrey conditionalization Jill will update her beliefs, in light of Ian's evidence, using the following formula:
\[P_f(H) = P_i(H|E)\cdot P_f(E) + P_i(H|\sim E) \cdot P_f(\sim E)\]
This equation says that Jill's final belief in the hypothesis, $P_f(H)$, is equal to her credence that the evidence is real, $P_f(E)$, multiplied by the belief she would obtain via strict conditionalization on that evidence, $P_i(H|E)$, plus her credence that the evidence did not occur, $P_f(\sim E)$, multiplied by the belief she would have by strict conditionalization if it had not occurred, $P_i(H|\sim E)$.

The Jeffrey conditionalization formula alone is not sufficient to fix Jill's belief; we also need to specify the credence that Jill assigns to the evidence, which we assume to be conditional on its source.  We consider two ways of doing this.  In both cases, we take Jill's credence to be a decreasing function of the difference in beliefs between the two agents.  On the first approach, we suppose that agents trust those with more similar beliefs, but that they completely ignore evidence from agents whose beliefs diverge by too much.  On the second approach, we assume that when Ian's beliefs are very far from Jill's, she actually is so suspicious of him him that she updates away from what his evidence seems to show.  (Likewise, \citet{cook2016rational} find that conservatives engage in contrary belief updating upon learning about the scientific consensus on climate change.)

The idea here is that as a responsible scientist, Jill must assess the reliability of the evidence produced by her colleagues.  There are many ways in which she might attempt to do this, but one plausible heuristic is to evaluate the reliability of the evidence on the basis of her perception of Ian's past epistemic success.  Jill's own beliefs are her clearest guide to evaluating whether Ian has succeeded in forming reliable beliefs, and so in cases where their beliefs differ, Jill supposes that Ian must be less reliable.

We will make the second approach, where evidence supporting theory B might actually decrease credence in that theory, precise first, as in some ways the formula is simpler.  Let $d$ be the absolute value of the difference between the Ian and Jill's credences.  We use the following to characterize the uncertainty that Jill assigns to evidence produced by Ian as a function of $d$:
\begin{equation}P_f(E)(d) = \max(\{1-d\cdot m \cdot (1-P_i(E)),0\}).\label{antiUpdate}\end{equation}
Here $P_i(E)$ is Jill's initial probability of the evidence occurring given her beliefs about theory A and theory B and $m$ is a multiplier that captures how quickly agents become uncertain about the evidence of their peers as their beliefs diverge.  Notice that $d$, the distance between beliefs, will vary between 1, if one agent has complete belief in theory B and the other in theory A, and 0, if both agents have the same credences in theory B. As $d$ approaches 0, $P_f(E)$ approaches 1, meaning that the agent thinks the evidence is more and more certain.  (Observe that every agent is distance $0$ from themselves, and so they treat their own evidence as certain.)   In this case, Jill's update rule approximates strict conditionalization very well.

As $d$ increases, meanwhile, at some point $d\cdot m = 1$.  (If $m = 2$, for example, this occurs when $d = .5$.)  At this point the certainty that Jill assigns to Ian's evidence is simply equal to the $P_i(E)$---i.e., to Jill's prior probability of the evidence occurring.  In other words, she completely ignores evidence from Ian, in the sense that her credence about theory B will be unchanged in light of the evidence from Ian.  As $d$ further increases, so that $d\cdot m > 1$, $P_f(E)$ becomes smaller than the prior belief that the evidence would occur, $P_i(E)$.  In other words, Jill believes that the evidence is less likely than she would have otherwise simply because Ian shared it.  Finally, since $P_f(E)$ is to be interpreted as an agent's credence, we require that $P_f \geq 0$.

For the approach where agents simply ignore evidence gathered by researchers they do not trust, we use the following alternative alternative function to describe Jill's uncertainty about Ian's evidence as a function of $d$:

\begin{equation}\label{noAntiUpdate}
P_f(E)(d) = 1-\text{min}(1, d\cdot m) \cdot (1-Pi(E)).\end{equation}

In this formula, we do not let $d\cdot m$ grow larger than 1.  This means that as beliefs diverge, there is a point, $d\cdot m = 1$, after which agents simply ignore the data of their peers, always assigning $P_f(E)=P_i(E)$.  The multiplier $m$ then determines how far apart beliefs have to be before individuals ignore the evidence of another researcher.

\begin{figure}
\centering
\includegraphics[width=.8\textwidth]{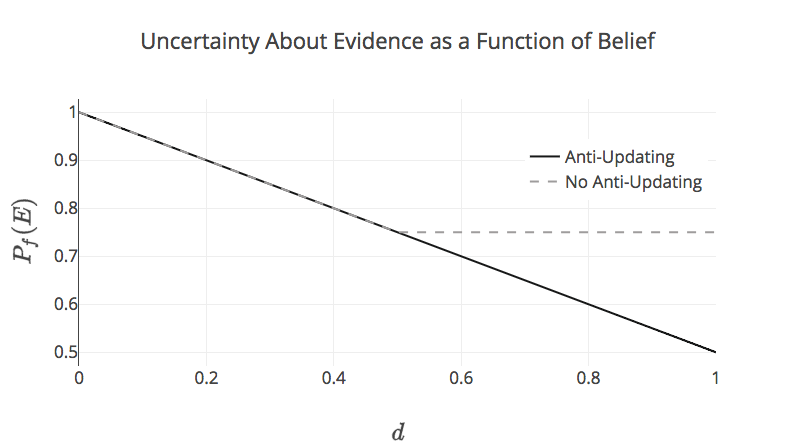}
\caption{An agent's uncertainty about evidence as a function of distance between credences both for anti-updating and simply ignoring evidence, $m = 2$ and $P_i(E) = .75$.}
\label{fig:uncertainty}
\end{figure}

In what follows, we will refer to the approach using Eq.~\eqref{antiUpdate} as the one with ``anti-updating''; the approach using Eq.~\eqref{noAntiUpdate} will be the one with no anti-updating. Figure \ref{fig:uncertainty} shows an example of what each of these functions would look like for an agent with prior probability $P_i(E) = .75$ and $m = 2$.  The x-axis tracks distance in belief between the two agents, which, as noted, ranges from 0 to 1.  The y-axis tracks agent certainty in evidence as a function of this distance.  Up to a distance $d = .5$, the two functions are the same, and certainty in evidence decreases linearly in distance between beliefs. After that, the anti-updater thinks it is less likely the evidence occurred than her prior would suggest, and the agent who ignores data thinks the evidence is just as likely as her prior belief predicts.

Of course \eqref{antiUpdate} and \eqref{noAntiUpdate} are arbitrary.  The results we describe will be qualitatively similar using different functions, as long as $P_f(E)$ decreases sufficiently quickly with $d$.  Indeed, as a robustness check, we ran simulations for several additional functions for $P_f(E)$, including a logistic function (eg., $P_f(E)(d) = 1/(1+\exp(m*(d-1/2)))$) and an exponential (eg. $P_f(E)(d) = \exp(-m*d)$).\footnote{Observe that the values of $m$ that make sense to consider vary between these functions and the linear functions we focus on; for instance, for the logistic function, we studied $m=5,7.5,10,12.5,15,17.5,20$, and for the exponential we looked at $m=1,2,3,4,5,6,7$.}  We found that the results were qualitatively robust across different functional dependencies, as long as there is some value $d_0$ such that for $d>d_0$, $P_F(E)(d) \leq P_i(E)$.  If this condition is not met, one does not get stable polarization, because even as $d$ approaches 1, agents continue to exert some influence on one another.\footnote{So, for instance, if one considers a function of the form $P_f(E)(d) = (1-P_I(E))/(1+\exp(m*(d-1/2)))+P_i(E)$, which is bounded from below by $P_i(E)$ and never achieves this value on $d\in[0,1]$, polarization is not stable.  However, this sort of exponential drop-off in influence as a function of $d$ dramatically increases converge times, and so we find that polarization may still be effectively stable, a result that amplifies the arguments we give below.}$^{,}$\footnote{There is another way of doing all of this, which is to suppose some probability distribution that describes Jill's credences about Ian's dispositions to share $E$ given that $E$ did and did not obtain, given Jill's own prior $P_i(E)$ and $d$, and then have her use Bayes' rule to find her posterior $P_f(E)$, given that Ian reports $E$.  But observe that doing this in detail would require an enormous number of modeling choices that would also be largely arbitrary, and at the end of the day, one would find a formula with the salient features of \eqref{antiUpdate} and \eqref{noAntiUpdate} (i.e., a monotonically decreasing function in $d$ whose range lies in the relevant interval).  More, one can always use Bayes' rule to work backward from Eq. \eqref{antiUpdate} or Eq. \eqref{noAntiUpdate} to a relationship between the conditional probabilities $P(\text{Ian shared E}|E)$ and $P(\text{Ian shared E}|\sim E)$ that must hold if we assume that Jill had such credences and that she arrives at $P_f(E)$ via strict conditionalization.  And so these formulae can themselves be interpreted as reflecting precisely the results of this procedure for (families of) distributions that might represent Jill's beliefs about Ian's dispositions.}

\section{Results}
\label{sec:results}

Our model involves several parameters.  First, we vary the size of the scientific community, testing values for $K$ ranging from 2 to 20.  We also vary the difference in success rate between the well-understood action A and the better action B.  The probability that A succeeds is always $P_A = .5$, and for the probability that B pays off we considered $P_B$ ranging from $.501$ to $.8$.  This parameter controls how equivocal evidence will tend to be.  As $P_B$ approaches $P_A$, the two actions become increasingly difficult to discriminate, and the number of spurious results, i.e., results suggesting that action A is actually preferable to action B, increases.  We vary the number of tests each scientists runs every round, $n$ from 1  to 100.  This parameter will also influence, on average, how many spurious results occur and how much they affect belief.  Smaller values of $n$ are analogous to lower powered studies which are more likely to yield misleading results.

Finally, we vary $m$, the multiplier that determines how quickly scientists begin to mistrust those with different beliefs, looking at values from 0 to 3.\footnote{Increaing $m$ beyond this range had little effect on the results since trust already drops off steeply when $m = 3$.}  When $m = 0$, agents do not discount evidence based on belief at all.  When $m = 1$, the agents never fully discount the evidence of other scientists (or engage in anti-updating), though they become less trusting of the data as beliefs diverge.  When $m$ is higher, agents completely ignore or else anti-update on individuals whose beliefs are more than some distance away from their own. (For instance, as demonstrated in figure \ref{fig:uncertainty}, if $m = 2$, this threshold is .5.)  One thing we do not vary is the network structure of the model.  We test only complete networks, meaning that agents communicate with every other member of their community.  This means that when agents polarize, it is in spite of the fact that they receive data from all their peers, and is not a result of differential access to information.\footnote{Note that this means that distance in belief may be reconceptualized as a weight on each edge of the network, so that initially there are random weights assigned, and then over time the network evolves so that some connections become stronger and others weaker.  In this sense, the model can be conceptualized as a dynamic network.}

For each combination of parameter values we ran 1k trials of simulation, until one of three measurable outcomes was reached---communities arrived at correct consensus (all beliefs were greater than $.99$), incorrect consensus (all beliefs were less than $.5$), or polarization.  (We will say more about just what `polarization' involves shortly.)

\subsection{Ignoring Data}

Let us first consider the version of the model where agents ignore data shared by those with very different beliefs.  Under many circumstances we observed stable polarization in these scientific communities.  In this version of the model, polarization involves the emergence of two subgroups, one whose members all have credence $>.99$, and the rest with a variety of stable, low credences, such that they prefer the worse theory.  (More precisely, a stable outcome is one in which every agent either (a) has credence $>.99$ or else (b) has credence $<=.5$ such that their distance to all agents whose credence is $>.99$ satisfies $m*d>=1$.)  Because the agents with low credences are outside the ``realm of influence'' of those testing the informative theory, they do not update their beliefs.\footnote{Notice that this operationalization of polarization means that simulations where one individual holds a stable minority opinion still counts as polarization.  One might object that true cases of polarization will involve more evenly sized subgroups.  For practical reasons, we prefer not to choose an arbitrary cut-off for what proportion of a population must hold each opinion in order to count as truly polarized.  \citet{bramson2017understanding} discuss subtleties of how groups can polarize.}

To reiterate, when this model is run without uncertainty based on divergence of beliefs (when $m = 0$), every simulation will arrive at full community consensus usually on the correct theory, and sometimes on the wrong one.\footnote{The probability of correct versus incorrect convergence varies based on parameter values.  See \citet{zollman2007communication,zollman2010epistemic,rosenstock2017epistemic} for more.}  In our models, on the other hand, over all parameter values, we found that only $10\%$ of trials ended in false consensus, $40\%$ in true consensus, and $50\%$ in polarization.  These values should not be taken too seriously, since parameter choices influence where and when polarization happens, but the point is that adding evidential assessments based on shared belief dependably generates stable polarization.

\begin{figure}
\centering
\includegraphics[width=.8\textwidth]{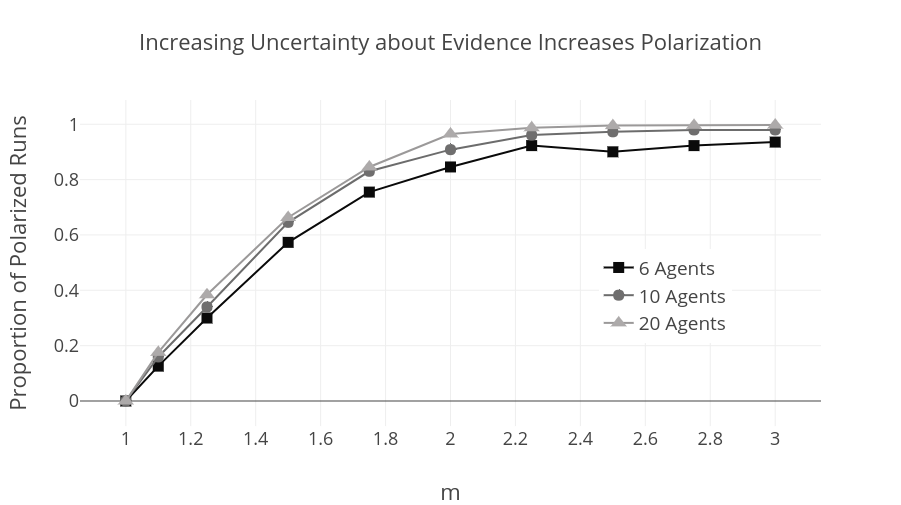}
\caption{Increasing $m$ increases probability of polarization. $n = 50$, $p_B = .7$.\label{fig:Polarm}}
\end{figure}

The multiplier determining how quickly scientists discount the evidence of others, $m$, strongly determines how often polarization occurs. Of course, this is no surprise since the mechanism necessary to generate polarization is strengthened and weakened via this multiplier.  Figure \ref{fig:Polarm} shows this effect for simulations with different numbers of agents.  This trend, and others reported in this section, are general across parameters unless otherwise noted.   As is clear, higher $m$ leads to higher polarization. As we also see in this figure, the chance of reaching a polarized outcome increases slightly in community size.  This is simply because with a larger group the chances are better that at least one agent stably disagrees with the rest.

\begin{figure}
\centering
\includegraphics[width=.8\textwidth]{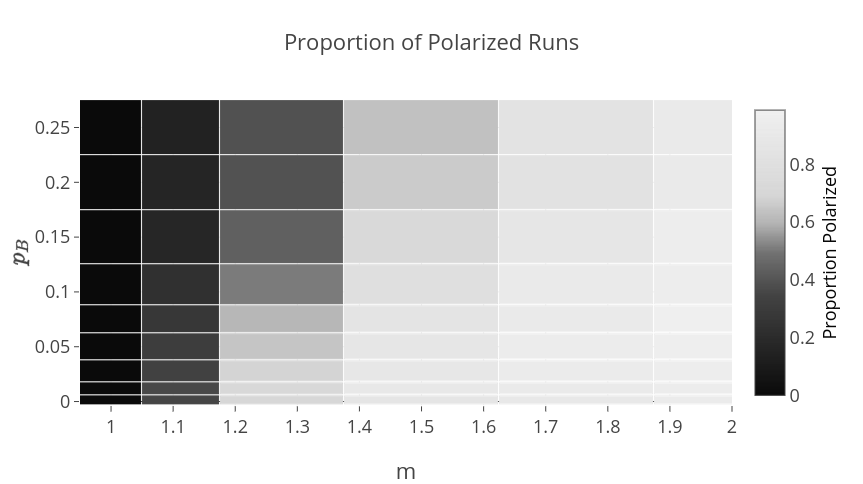}
\caption{Increasing $m$ increases probability of polarization, this is mitigated when $p_B$ is higher. $K = 10$, $n = 20$.\label{fig:polarheat}}
\end{figure}

Figure \ref{fig:polarheat} demonstrates this phenomenon in heatmap form for different values of $m$ and $p_B$ (success rates for the better theory).  As is evident again, higher $m$ means more polarization.  This effect is ameliorated somewhat when $p_B$ is higher, because strong evidence for theory B tends to drive beliefs up more quickly.  Still, even when the better theory is obviously better, once trust is low enough, polarization emerges in almost all cases. We see here the robustness of this effect across parameter values.

This sort of polarization will only occur when $m$ is high enough that there are credences an agent could hold and be entirely unaffected by evidence coming from the part of the community that converges to high credence in theory B.  For instance if $m = 1.1$, all agents update on the evidence of almost all other agents.  But, an agent with credence .04 will not update on evidence from an agent with credence .99 at all, meaning polarization is possible.  When $m = 1$, on the other hand, there are no stable polarized outcomes.  Eventually all agents will reach consensus on either theory A or theory B.  But even in this case, Jeffrey conditionalization can lead to transient polarization---the temporary existence of two subgroups one of which has very high credences ($>.99$) and the other with credences $<.5$.

\begin{figure}
\centering
\includegraphics[width=.8\textwidth]{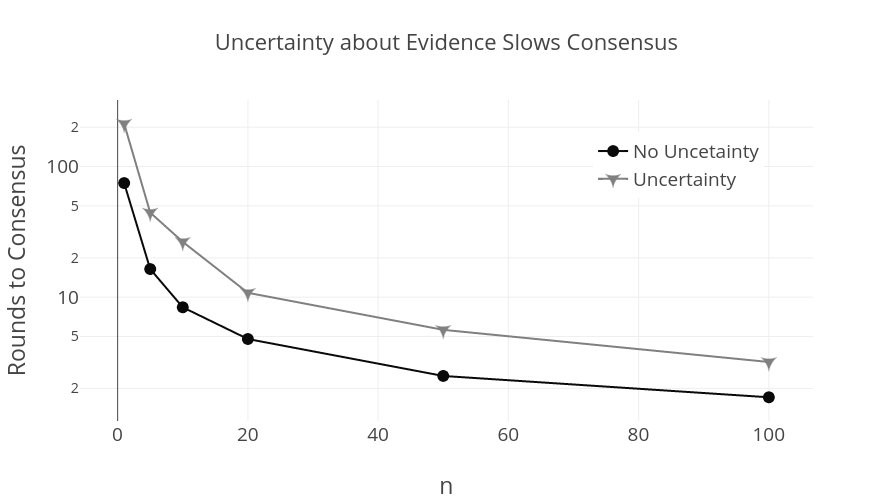}
\caption{Uncertainty ($m = 1$) slows consensus compared to no uncertainty ($m = 0$), even though polarization is not possible for either of these values of $m$. $p_B = .55$, $K = 6$.
\label{fig:Polarcon}}
\end{figure}

Moreover, although the whole population will always converge to some consensus in this case, the time to convergence is substantially longer than when $m=0$.  And in the meantime there is a potentially very long period during which some portion of the community is mistrustful of an emerging consensus.  Figure \ref{fig:Polarcon} shows the average speed at which a community reaches consensus when $m = 1$ or $m = 0$, for various numbers of pulls, $n$.  Notice that the y-axis is on a log scale to make the trend more clear.  For all values, adding uncertainty about the evidence of those with different beliefs slowed convergence to consensus by a factor of 2 or 3 on average.

This occurs because the addition of uncertainty about evidence and Jeffrey conditionalization to the model creates new updating dynamics.  The key, here, is that although all agents in our model have access to the same evidence every round of simulation (because they are on complete networks) they all treat that evidence differently.  Say an agent with credence .55 happens to gather a string of data spuriously supporting theory A, and another agent with credence .9 happens to gather a string of data that correctly supports theory B.  Despite all agents receiving the same data, those with already low credences will tend to decrease their credence further, while those with high credences will increase them in the same round.

This sort of behavior can lead to feedback loops, by which agents who have more similar credences gradually diverge---and then trust one another less as a result.  For instance, consider two agents with initial credences .6 and .3.  The .6 agent tests the informative action and generates results indicating that theory B is, in fact, better.  They update by increasing their credence in theory B.  The .3 agent also updates their credence in theory B, but by a much smaller amount because the distance between them (.3), leads her to discount the evidence.  Say that their new credences are .88 and .45.  The .88 agent tests the informative theory again.  This time, again, they both update in the same direction, but the distance between them is now .43, so the .45 agent is more mistrustful than before.  Via this sort of process, a belief gap and a trust gap can emerge between those who have become converts to a new theory and those who remain skeptical.\footnote{The values in this example were calculated assuming that $p_B = .6$, $n = 10$, and assuming that the .6 agent sees 7 successes in their test.}

There is one last result to mention, which is that the size of the multiplier determines the proportion of the community, on average, that ends up holding the false belief when polarization happens.  For large multipliers, and thus higher levels of mistrust, more agents tend to end up believing in the worse theory A.  If they are initially skeptical, they also do not trust those who take the informative action, and so tend to stick with their skeptical beliefs. For smaller multipliers, and so more trust in others, a larger percentage of agents have beliefs that are pulled up to the better theory by those testing this theory.  What this means is that, in general, mistrust in others with different beliefs results in a much higher degree of incorrect belief than would otherwise occur.

\begin{figure}
\centering
\includegraphics[width=.8\textwidth]{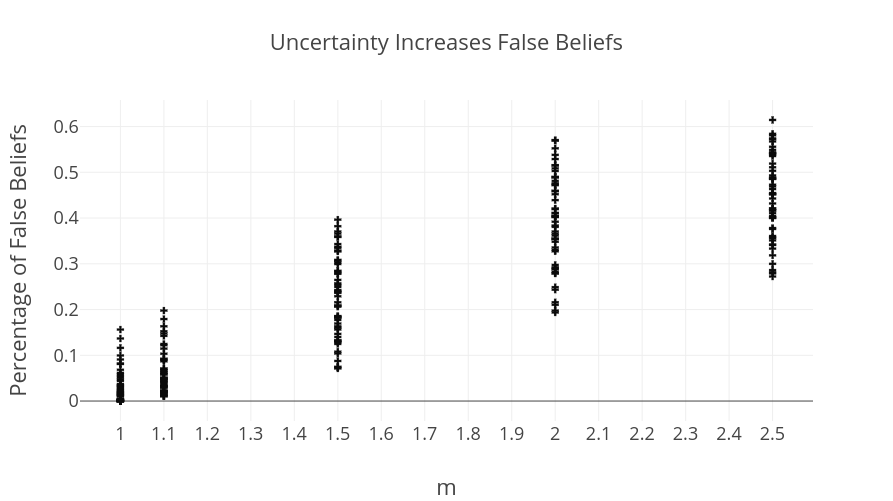}
\caption{Average percentage of false beliefs in scientific communities for different parameter values as a function of $m$, which tracks how quickly agents become uncertain in the evidence of others as beliefs diverge.\label{fig:Polarfal}}
\end{figure}

Figure \ref{fig:Polarfal} shows the average number of individuals who end up stably convinced in the incorrect theory as the multiplier increases.  Each data point in this figure is the percentage of false beliefs across runs of simulation for one set of parameter values.  As is clear, as the multiplier $m$ increases, the average percentage of false beliefs does too.  A lack of trust in others based simply on their beliefs leads to a community in a worse epistemic state.  Of course, this is in a model where all actors are epistemically reliable in the sense that they gather and share dependable data, and there are no biased agents in the scientific network.  In the conclusion, we will discuss when and why it might be a good thing not to trust those in a scientific network.

\subsection{Anti-updating}

We now turn to the case where agents are so mistrustful of those with different beliefs that they sometimes expect others to actively seek to mislead, and thus use Eq. \eqref{antiUpdate} to assign credences to evidence reported by others.  In this case, again, we find that communities reach stable polarization.  The actual outcomes are somewhat different with anti-updating.  Mistrust in those with high credences now drives the beliefs of those with low credences further and further down over time.  This is analogous to the conservative who, upon learning about scientific consensus about climate change, updates to greater skepticism about climate change \citep{cook2016rational}.  When polarization occurs in these models, then, the agents form two subgroups whose credences are either $>.99$, or $<.01$, and are increasingly unlikely to ever leave these ranges.

In models with anti-updating, polarization arises slightly more often than in models without it.  This occurs when there are individuals who might have been positively influenced by their own results, or those of nearby agents taking the informative action, but who are so mistrustful of those with high credences that their anti-updating overwhelms the good evidence reaching them.  In cases without anti-updating these individuals might eventually reach the correct belief, but in the presence of anti-updating they do not because the comparative zealots are too influential.

\begin{figure}
\centering
\includegraphics[width=.8\textwidth]{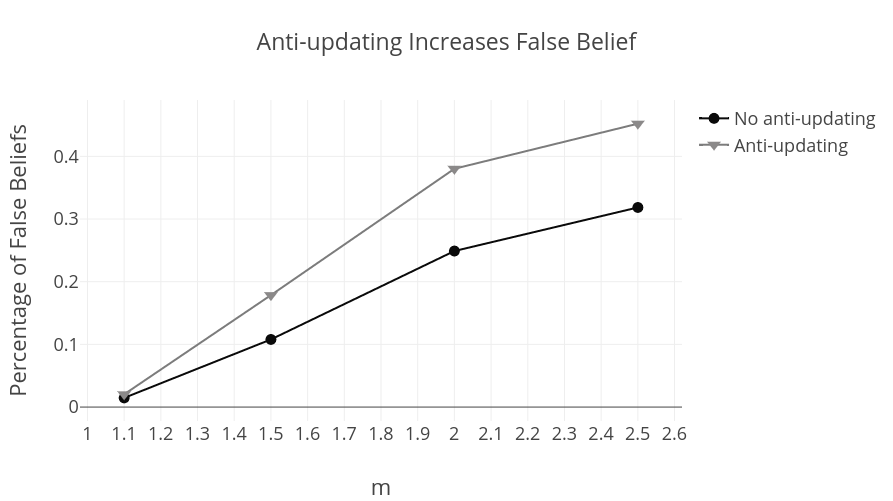}
\caption{Models where actors anti-update tend to have a larger portion of false beliefs, as a result of increased polarization. $p_B = .7$, $n = 10$, population size = 20.
\label{fig:polaranti}}
\end{figure}

Additionally, anti-updating tends to increase the number of individuals arriving at the incorrect belief in comparison to simply ignoring others' data.  Figure \ref{fig:polaranti} shows the percentage on average of agents arriving at true and false beliefs in the two types of model.\footnote{The significance of the difference between the anti-updating case and the ignoring case varies across parameter values.  In a few cases the community did slightly better on average in the anti-updating case, usually for small communities where results were more stochastic.}  As is obvious in this case, anti-updating leads to worse beliefs, and this is more dramatic as $m$ increases.  Anti-updating means that more individuals who might be convinced by other moderates like themselves end up driven to low credences.

\section{Conclusion}

\citet{mayo2011independence} describe what they call the \emph{Independence Thesis}, which consists in the claims that, ``rational individuals can form irrational groups, and, conversely, rational groups might be composed of irrational individuals'' (653).  They point out that the entire field of social epistemology is undergirded by the assumption that community and individual rationality come apart, meaning that in order to best understand the progress of knowledge, we need to focus on the group level rather than just on individual rationality.\footnote{As \citet{mayo2011independence} prove using network epistemology models similar to the ones we employ here, there are rules for exploration in such models that are ideal for the individual, but not the group, and vice versa.  Other formal work in social epistemology focuses on this idea as well.  Both \citet{kitcher1990division} and \citet{strevens2003role}, for example, explore how to generate an ideal division of cognitive labor in science despite the individual rationality of always working on the most promising theory.}

The models here fall broadly under the heading of showing how what makes sense for an individual, and what makes sense for a group, can come apart.  Clearly, treating the evidence of those with different beliefs as uncertain can have detrimental effects from the group-level standpoint.  The greater $m$ is, the worse the average beliefs of the scientists in our models.  And when actors anti-update, this situation is exacerbated.\footnote{Notice that we do not discuss here potential benefits from transient polarization.  For example, \citet{zollman2010epistemic} argues for the importance of transient diversity of opinions in epistemic groups.  (Without this diversity, there is less chance that scientists spend enough time testing every plausible theory to see which is best.)  Since polarization ensures an extended diversity of beliefs, it may increase the chances that the scientific communities as a whole gathers good evidence about all plausible theories.  Likewise, in \citet{zollman2010epistemic}, a community can benefit from the presence of individuals with strong priors, who keep exploring a theory even when it looks unpromising. The problem, in his models and in ours, is individuals who are too stubborn, or who never update in light of untrusted evidence.  We also do not discuss potential benefits of political polarization identified by political scientists, such as a more robust, argumentative discourse.  (See \citet{abramowitz2010disappearing} for a discussion.)}

On the other hand, while we might not want to label it as rational with a capital ``R'', there is something reasonable about deciding on an individual level whose evidence to trust on the basis of their currently held beliefs.  Suppose your Uncle Matt tells you that Hillary Clinton personally had 46 journalists killed, and that he has the documents to prove it.  If you also know Uncle Matt believes that your aura will be more aligned when there is lots of quartz in the ground, you might take his documents less seriously, and with good reason.  If your pediatrician tells you that cow's milk has ``no nutrition in it'', it is, again, reasonable not to trust other data she might later share.  In the case of scientists, those who directly uptake evidence without considering the source might even be considered epistemically irresponsible.  Most scientists do not trust the evidence of known ``quacks'', and for good reason.

Nonetheless, one possible take-away from the models presented here might be that in a scientific community we should do whatever it can to drop this heuristic, and evaluate all evidence in the same way.  But notice that these models do not capture the type of situation in which discounting the evidence of others makes sense. In some cases, individuals in a scientific community intentionally mislead peers for their own benefit.  In 1954, for example, the Tobacco Industry Research Committee was created with the stated goal of investigating the health effects of smoking.  In fact the committee was a propaganda machine created by the heads of major tobacco firms, but from the point of view of those receiving information from them, there was little to distinguish this source from others.\footnote{This history is drawn from \citet{Oreskes+Conway}, who document in great detail the work done by big tobacco to obscure the emerging consensus over the health dangers of smoking.  See also  \citet{holman2015problem}, \citet{OConnor+WeatherallBook}, and \citet{Weatherall+etal}.}  Among their activities was selectively sharing intentionally misleading results, with the intention of manipulating beliefs.  In such cases, one would certainly prefer to evaluate the evidence they share differently from evidence shared by unbiased scientists.

\citet{holman2015problem} investigate a network epistemology model much like the one we look at here, but which includes an ``intransigently biased agent''.  The biased agent only tests the worse theory, and when they do so, their probability of success is artificially inflated. Holman and Bruner find that such an actor tends to influence the beliefs of their community in a negative way, but that if other scientists have an option to devalue their evidence the problem is ameliorated.  Their model incorporates this devaluation by placing weights on every network edge.  When a scientist receives new evidence, they do a t-test based on their current credence in the theory.  If the evidence seems particularly unlikely given their beliefs, they reduce their weight on that edge.  As they show, ``The problem posed by intransigently biased agents can be alleviated if agents learn to identify and trust good informants." (966).  In other words, in the model with industry actors, the option to ignore others' data is crucial to the success of the community.

Notably, ignoring in their model is based on a match between \emph{evidence} and one's own credence, not between another's credence and one's own as in our model.  As they note some scientists sometimes end up in the biased agent's sphere of influence because their credences are influenced by the biased agent's evidence, and then the biased evidence looks plausible to them.  This, in itself, is a type of polarization where one part of the community holds a different belief and takes a different action from the other part, and the two groups have little influence over each other.  It results, as in other models of polarization, from the fact that there is a dependence between the beliefs of scientists and social influence of the biased agent.  But, despite the possibility of this sort of polarization, the ability to evaluate evidence based on whether it accords with one's scientifically informed belief is a good thing in these models.  It significantly improves the epistemic states of scientists.

Additionally, one might want to distrust evidence from scientists who are less dependable than others, which, again, is a possibility our models does not address. \citet{barrett2017self} present a model that approximately captures this sort of situation.  They consider a networked group of agents who all have an option to test the world, with different characteristic rates of success.  These agents can also choose to consult the conclusions of their peers, again with different rates of successful social transmission.  They find that if they allow such networks to evolve, i.e. the agents update their probabilities of testing the world and consulting other agents based on the success of these strategies, the final results are often very successful compared to random starting points.  In other words, the ability to choose to listen to those who have been epistemically reliable in the past helps all agents develop better beliefs.  Again, we see a case where it is good not just for the individual, but also for the group, for agents to ignore the evidence of some peers and favor the evidence of others.

It seems, then, that while there are multiple heuristics available for treating evidence from unreliable or biased agents, any of which may seem justified at the individual level, they can lead to different outcomes at the group level.  In our model, scientists condition their trust in evidence of others based on distance in belief, which is simple but can have bad effects for the community at large.  The heuristic in \citet{holman2015problem} involves updating less strongly on evidence that does not fit with one's current beliefs, and can also lead to polarization.  But consider a different---albeit more difficult---heuristic: suppose that scientists learn to be uncertain about sources whose evidence persistently differs, statistically, from most other sources.  These scientists can also avoid being misled, and may do so without negatively affecting the epistemic performance of the community.  The point is that while belief similarity and confirmation bias are easy heuristics to depend on in deciding who to trust, there are other ways to make this decision that do not risk driving a community towards polarization.

We will conclude with a discussion about what the models we have presented here can and cannot tell us about polarization in real scientific communities, returning to the case of chronic Lyme disease.  Obviously these models are highly idealized.  Real humans, for example, are not perfect Bayesians, and many aspects go into scientists' decisions about what data to trust.  Nonetheless, the models can do a few things.  First, they show how, in principle, a situation like that in the chronic Lyme case can arise.  We do not need to suppose that anyone is a bad researcher (in our models all agents are identical), or that they are bought by industry, or even that they engage in something like confirmation bias or other forms of motivated reasoning to see communities with stable scientific polarization emerge.  All it takes is some mistrust in the data of those who hold different beliefs to get scientific polarization.  In addition, these models provide a robustness check on previous models of polarization by showing once more how the general feature responsible for it---dependence between shared beliefs/opinions/features and social influence---can lead to polarization even in a situation where it might not be expected because there are clear reasons to prefer one belief over another, and all agents have the capacity to directly test their beliefs.

The models also suggest a few interventions if, indeed, mistrust of those with different opinions is helping to drive polarization in the chronic Lyme case.  In particular, one possible solution is to find a neutral party---for example, a group of independent researchers convened through the National Science Foundation---to do meta-analyses and survey articles on Lyme disease.  The hope is that entrenched researchers suspicious of the `other side' might nonetheless be willing to trust individuals with beliefs less divergent from their own.

\section*{Acknowledgments}

Thanks to Justin P. Bruner, Calvin Cochran, and the School of Philosophy at Australian National University where most of the research for the paper was
carried out.  This material is based upon work supported by the National Science Foundation under grant no. STS-1535139.

\bibliographystyle{elsarticle-harv}
\bibliography{Polarizationbib}

\end{document}